\documentclass[11pt]{article}
\usepackage{amsmath,amsfonts,amssymb,amsbsy,amscd,amsgen}
\usepackage{amsthm}  
\usepackage{color}
\usepackage{appendix}
\usepackage[]{hyperref}  
\usepackage{authblk}
\usepackage{float}
\usepackage{subfig}
\usepackage{graphicx}
\graphicspath{
{Fig/}
}

\newcommand{\id}{\mathrm d}

\newcommand{\tr}{\mbox{tr}}

\newcommand{\pard}[2]{\frac{\partial #1}{\partial #2}}
\newcommand{\SDiff}{\mbox{SDiff}}
\newcommand{\Diff}{\mbox{Diff}}

\begin{document}
\title{Variational Lagrangian formulation
of the Euler equations for incompressible flow: A simple derivation}
\author[1]{Mohammad Farazmand}
\author[2]{Mattia Serra}
\affil[1]{{\small Department of Mechanical Engineering, Massachusetts Institute of Technology, 
77 Massachusetts Avenue, Cambridge, MA 02139–4307, USA}}
\affil[2]{{\small School of Engineering and Applied Sciences, Harvard University,
29 Oxford Street, Cambridge, MA, 02138}}
\date{}

\maketitle

\begin{abstract}	
	In 1966, Arnold~\cite{arnold1966} showed that the Lagrangian flow of ideal incompressible fluids (described by Euler equations) coincide with the geodesic flow on the manifold of volume preserving diffeomorphisms of the fluid domain. Arnold's proof and the subsequent work on this topic rely heavily on the properties of Lie groups and Lie algebras which remain unfamiliar to most fluid dynamicists. In this note, we provide a simple derivation of Arnold's result which only uses the classical methods of calculus of variations. In particular, we show that the Lagrangian flow maps generated by the solutions of the incompressible Euler equations coincide 
    with the stationary curves of an appropriate energy functional
    when the extremization is carried out over the set of volume-preserving diffeomorphisms. 
\end{abstract}

\section{Introduction}
Euler equations for the motion of an incompressible fluid over a domain $D\subset\mathbb R^d$
read
\begin{subequations}\label{eq:euler}
\begin{equation}
\partial_t v + v \cdot \nabla_x v +\nabla_x p =0,
\end{equation}
\begin{equation}
\nabla_x \cdot v = 0,
\end{equation}
\begin{equation}
v(0,x)=v_0(x),
\end{equation}
\end{subequations}
where the vector field $v:[0,T]\times D\to\mathbb R^d, (t,x)\mapsto v(t,x)$ denotes the fluid velocity field and
the scalar field
$p:[0,T]\times D\to\mathbb R, (t,x)\mapsto p(t,x)$ denotes the fluid pressure. 
Here, we consider two-dimensional ($d=2$) and three-dimensional ($d=3$) flows.
The Euler equations must be supplied with the appropriate boundary conditions on the 
boundary $\partial D$ of the domain $D$. Here, we assume that the boundary is
smooth and impose the no-flux boundary condition
\begin{equation}\label{eq:BCs}
v(t,x)\cdot n(x)=0,\quad \forall (t,x)\in[0,T]\times \partial D,
\end{equation}
where $n(x)$ denotes the unit vector normal to the boundary at the point $x$.
We assume that unique solutions (in the strong sense) to the initial value problem~\eqref{eq:euler}
exist over the time interval $[0,T]$ and that these solutions are sufficiently smooth.

Fluid particle trajectories satisfy
\begin{equation}
\dot g(t,a) = v(t,g(t,a)),\quad \forall a\in D,
\label{eq:ode1}
\end{equation}
where $g(t,a)$ denotes a trajectory passing through the point $a\in D$ at the initial time $t=0$, so that $g(0,a)=a$ (cf. figure \ref{fig:VectorField}).
Since it defines the Lagrangian fluid trajectories, we refer to $g$ as the \emph{Lagrangian flow map}.
If the velocity field $v(t,x)$ is sufficiently smooth, then the maps $g(t,\cdot):D\to D$
are diffeomorphisms of the domain $D$, i.e., $g(t,\cdot)\in\Diff(D)$. Here, $\Diff(D)$
denotes the set of all diffeomorphisms of the set $D$. 

We recall two well-known facts:
\begin{enumerate}
\item Since the flow is incompressible, we have 
\begin{equation}
\det [\nabla_a g(t,a)]=1,
\label{eq:volpres}
\end{equation}
for all $t\in[0,T]$ and $a\in D$ (see, e.g., Ref.~\cite{arnold78}).
\item If $v(t,x)$ is a solution of the Euler equations~\eqref{eq:euler}, 
taking the time derivative of equation~\eqref{eq:ode1} leads to
\begin{equation}
\ddot g (t,a) + \nabla_x p(t,g(t,a))=0,\quad g(0,a)=a,\quad \dot g(0,a)=v_0(a),
\label{eq:ddotg1}
\end{equation}
where $p$ and $v_0$ are the pressure and the initial velocity that appear in the Euler equations.

The converse is also true: Assume a one-parameter family of 
volume-preserving diffeomorphisms $g(t,\cdot)$
satisfy~\eqref{eq:ddotg1} for a given function $p(t,x)$. Defining $v(t,g(t,a))=\dot g(t,a)$, it is not hard to 
show that the pair $(v,p)$ satisfies the Euler equations~\eqref{eq:euler}.
\end{enumerate}

Property~\eqref{eq:volpres} implies that the Lagrangian flow maps $g(t,\cdot)$
belong to the infinite-dimensional manifold of volume-preserving diffeomorphisms, i.e., 
$g(t,\cdot)\in\SDiff(D)\subset \Diff(D)$ (cf. figure \ref{fig:Geodesic}). Here, $\SDiff(D)$
denotes the set of all volume-preserving diffeomorphisms of the set $D$.

Arnold~\cite{arnold1966} showed that the Lagrangian flow maps arising from
the ODE~\eqref{eq:ddotg1} are in fact geodesics on the manifold $\SDiff(D)$. Arnold's proof
and the subsequent work on this topic make heavy use of the geometry of
Lie groups and the notions of Lie algebras and Lie derivatives. 
The aim of this note is to derive Arnold's result using a simple approach that only 
relies on classical methods of calculus of variations that are familiar to most fluid dynamicists. To this end, 
we will skip many of the technical questions such as existence of extremizers and regularity of 
solutions (treated in detail in, e.g., Refs.~\cite{arnold1966,khesin2008,figalli2012}), and confine this 
note to a formal derivation of the results. 

\section{Constrained variational problem}
\begin{figure}[t]
	\subfloat[]{\includegraphics[height=0.38\columnwidth]{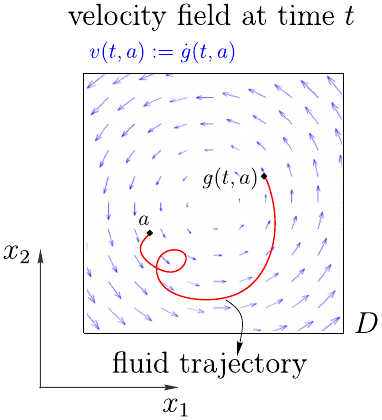}\label{fig:VectorField}}
	\hspace{.05\textwidth}
	\subfloat[]{\includegraphics[height=0.38\columnwidth]{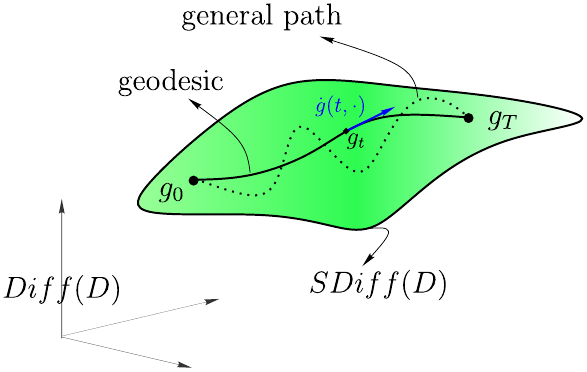}\label{fig:Geodesic}}
	\caption{(a) Illustration of $\dot g(t,\cdot)$ as a vector field over the physical fluid domain $D$ in the case of planar flows ($d=2$). The red curve represents a fluid trajectory with initial position $a$ and time-$t$ position $g(t,a)$.
(b) Sketch of the geodesic curve (solid black) connecting $g_0$ and $g_T$ on the manifold $\SDiff(D)$ shown in green. The dotted curve represents a general path connecting $g_0$ and $g_T$, and the blue arrow illustrates the time derivative $\dot g(t,\cdot)$ of $g$ at time $t$. }	
	\label{fig:StetchGeodesicandFlow}
\end{figure}

The set of all volume preserving diffeomorphisms of $D$, i.e. $\SDiff(D)$, 
turns out to have the structure of a manifold~\cite{khesin2008} (in fact, $\SDiff(D)$ is an infinite-dimensional Lie group). 
As such, it is interesting 
to investigate its geodesic curves. Consider two elements $g_0$ and $g_T$
of $\SDiff(D)$.
A geodesic curve connecting these two `points' is the curve whose endpoints are
$g_0$ and $g_T$ and whose length is minimal among all smooth curves with the same endpoints (see figure \ref{fig:Geodesic}).
More precisely, geodesic curves minimize the length
\begin{equation}
\int_0^T \|\dot g(t,\cdot)\|\id t,
\label{eq:length}
\end{equation}
where $g(t,\cdot)$ is a smooth one-parameter family (i.e. a curve) of diffeomorphisms in $\SDiff(D)$
with $g(0,\cdot)=g_0$ and $g(T,\cdot)=g_T$.
The parameter $t$ will turn out to be the non-dimensionalized time in the Euler equations. 
The velocity of the curve is denoted by $\dot g(t,\cdot)$ which is the derivative with respect to the parameter $t$ (cf. figure \ref{fig:Geodesic}). 
This velocity evaluated over the fluid flow domain $D$ corresponds to the time-$t$ snapshot of the fluid velocity field as illustrated in figure \ref{fig:VectorField}.
Note that although one usually assumes the initial point 
$g_0\in\SDiff(D)$ to be arbitrary, in the context of fluid dynamics $g_0$ is the identity map since $g(0,a)=a$ for all $a\in D$.

For the norm $\|\cdot\|$, here we choose the $L^2$ norm on $D$ so that 
$$\|\dot g(t,\cdot)\|^2 = \int_D |\dot g(t,a)|^2\id a,$$
where $|\cdot|$ denotes the standard Euclidean norm on $\mathbb R^d$.
In order to find the geodesic curve, we need to minimize the length functional~\eqref{eq:length}
over all curves in $\SDiff(D)$ with the endpoints $g_0$ and $g_T$.
Carrying out this minimization leads to the geodesic equations whose solutions are
geodesics on $\SDiff(D)$.
Arnold~\cite{arnold1966} showed that this geodesic equation
is equivalent to the Euler equations for ideal incompressible fluids. 

In this note, instead of minimizing the length functional~\eqref{eq:length}, 
we minimize the energy functional
\begin{equation}
\mathcal E(g):=\int_0^T\frac12 \|\dot g(t,\cdot)\|^2\id t=\int_{0}^{T}\int_D \frac12 |\dot g(t,a)|^2\id a\id t,
\end{equation}
and show that the minimizers of this energy functional satisfy the Euler equations. 
Also, instead of directly minimizing the energy functional $\mathcal E$ over the volume-preserving diffeomorphisms 
$\SDiff(D)$, we formulate an equivalent constrained optimization problem
where we minimize the energy functional over all diffeomorphisms $\Diff(D)$ with the volume-preserving property 
enforced as an additional constraint. 
This problem can be written concisely as
\begin{subequations}\label{eq:opt_const}
\begin{equation}
\min_{g(t,\cdot)\in\Diff(D)} \mathcal E(g),
\label{eq:opt_const1}
\end{equation}
\begin{equation}
g(0,a)=g_0(a),\quad g(T,a)=g_T(a),\quad \forall a\in D
\label{eq:opt_const2}
\end{equation}
\begin{equation}
\det[\nabla_a g(t,a)] = 1,\quad \forall t\in[0,T],\quad\forall a\in D.
\label{eq:const_vol}
\end{equation}
\end{subequations}
Equation~\eqref{eq:opt_const1} is the minimization over all diffeomorphisms of $D$. 
The constraint~\eqref{eq:opt_const2} enforces the fact that the endpoints of the curve $g(t,\cdot)$
are prescribed and fixed. Finally, constraint~\eqref{eq:const_vol} requires that $g(t,\cdot)$ are volume preserving for 
all $t$ and therefore they belong to $\SDiff(D)$. 

In the next section, we solve the constrained optimization problem~\eqref{eq:opt_const} and show that
its solutions coincide with the solutions of the Euler equations~\eqref{eq:euler}.

\section{Solving the variational problem}
First, we define the augmented energy functional
\begin{equation}
\tilde{\mathcal E}(g,q) = \int_{0}^{T}\int_D \frac12 |\dot g(t,a)|^2\id a\id t+
\int_{0}^{T}\int_D q(t,a)\left(\det[\nabla_a g(t,a)]-1\right)\id a\id t,
\end{equation}
where $q:[0,T]\times D\to\mathbb R$ is a Lagrange multiplier. Note that the variation of the functional with respect to $q$,
evaluated at the extremizers, gives the volume-preservation constraint~\eqref{eq:const_vol},
\begin{equation}
\delta_q\tilde{\mathcal E}=0 \iff \det[\nabla_a g(t,a)]-1=0. 
\end{equation}

Next, we compute the first variation of the augmented energy functional with respect to the diffeomorphisms $g$.
To this end, we study the variations of $\tilde{\mathcal E}$ with respect to the perturbations $g+\epsilon h$
where $0<\epsilon\ll1$ and $h:[0,T]\times D\to\mathbb R^d$. Since the endpoints $g(0,\cdot)$ and $g(T,\cdot)$ are fixed, 
we require the perturbations to vanish at the end point, i.e., $h(0,\cdot)=h(T,\cdot)=0$. Furthermore, 
since the boundary of the domain has to remain invariant, we require the perturbations
to also vanish along the boundary for all times, i.e., $h(t,a)=0$ for all $t\in[0,T]$ and $a\in\partial D$.
In summary, $h$ has to satisfy
\begin{subequations}\label{eq:h_bc}
\begin{equation}
h(0,a)=h(T,a)=0,\quad \forall a\in D,
\label{eq:h_bc_time}
\end{equation}
\begin{equation}
h(t,a)=0,\quad \forall t\in[0,T],\quad \forall a\in\partial D.
\label{eq:h_bc_space}
\end{equation}
\end{subequations}
Note that the requirement~\eqref{eq:h_bc_space} is not a necessary condition for the perturbations $g+\epsilon h$
to leave the boundary $\partial D$ invariant. However, we make this assumption in order to simplify the following analysis.

For the perturbed energy functional $\tilde{\mathcal E}$, we obtain
\begin{align}
\tilde{\mathcal E}(g+\epsilon h,q) = &\mathcal E(g) + 
\epsilon\int_{0}^{T}\int_D \langle\dot g ,\dot h\rangle \id a \id t\nonumber\\
& +\int_{0}^{T}\int_D q\left\{\det[\nabla_a g +\epsilon\nabla_a h]-1 \right\}\id a\id t+\mathcal O(\epsilon^2)\nonumber\\
=&\mathcal E(g) -\epsilon \int_{0}^{T}\int_D \langle\ddot g , h\rangle \id a \id t\nonumber\\
& +\int_{0}^{T}\int_D q \left\{ \det[\nabla_a g]\det[I+\epsilon(\nabla_ag)^{-1}\nabla_a h]-1  \right\}\id a\id t+\mathcal O(\epsilon^2)\nonumber\\
= & \tilde{\mathcal E}(g,q) -\epsilon \int_{0}^{T}\int_D \langle\ddot g , h\rangle \id a \id t\nonumber\\
& +\epsilon\int_{0}^{T}\int_D q\;  \tr[(\nabla_ag)^{-1}\nabla_a h]\id a\id t+\mathcal O(\epsilon^2),
\label{eq:Eeps1}
\end{align}
where $\langle\cdot,\cdot\rangle$ denotes the Euclidean inner product. 
To obtain the above identities, 
we used integration by parts in $t$ and the well-known formula
\begin{equation}
\det[ I+\epsilon A] = 1+\epsilon\,\tr A+\mathcal O(\epsilon^2).
\end{equation}
Note that the boundary terms from the integration by parts vanish because of
property~\eqref{eq:h_bc_time}.

Let us define $G(t,a):=[\left(\nabla_a g(t,a)\right)^{-1}]^{\dagger}$ where $\dagger$ denotes the matrix transposition. 
The tensor $G$ satisfies the identity
\begin{equation}
\nabla_a\cdot[G^\dagger h] = \langle\nabla_a\cdot G, h\rangle + \tr[G^\dagger \nabla_a h].
\label{eq:G_id}
\end{equation}
We prove this identity in Appendix~\ref{app:G_id}.
It follows from a straightforward (but tedious) calculation that $\nabla_a\cdot G = 0$.
This observation, identity~\eqref{eq:G_id} and equation~\eqref{eq:Eeps1} together yield
\begin{equation}
\tilde{\mathcal E}(g+\epsilon h,q) = \tilde{\mathcal E}(g,q)
+\epsilon \int_{0}^{T} \int_D\left\{-\langle \ddot g,h\rangle+ q\, \nabla_a\cdot[G^\dagger h] \right\}\id a \id t+\mathcal O(\epsilon^2).
\label{eq:Eeps2}
\end{equation}
Using the identity $q\, \nabla_a\cdot[G^\dagger h]=\nabla_a\cdot[qG^\dagger h]-\langle \nabla_a q, G^\dagger h\rangle$, we rewrite the second term in the integral as
\begin{equation}
\begin{aligned}
\int_D q\, \nabla_a\cdot[G^\dagger h] \id a =& \int_D \left\{\nabla_a\cdot[qG^\dagger h]-\langle \nabla_a q, G^\dagger h\rangle \right\}\id a\\
=& \int_{\partial D} q\langle G^\dagger h, n\rangle\ dS-\int_D \langle \nabla_a q, G^\dagger h\rangle\ \id a
\label{eq:SpaceBC}
\end{aligned}
\end{equation}
where we used the divergence theorem
in the last identity above, and we recall that $n$ denotes the outward pointing unit vector orthogonal to the boundary $\partial D$. 
Furthermore, the boundary integral in equation~\eqref{eq:SpaceBC} vanishes due to property~\eqref{eq:h_bc_space}.

Combining equations~\eqref{eq:Eeps2} and~\eqref{eq:SpaceBC} we obtain
\begin{equation}
\tilde{\mathcal E}(g+\epsilon h,q) =\tilde{\mathcal E}(g,q) 
-\epsilon \int_{0}^{T} \int_D\langle \ddot g+G\nabla_a q,h\rangle\id a \id t + \mathcal O(\epsilon^2). 
\label{eq:Eeps3}
\end{equation}
Therefore, the first variation of the augmented energy functional with respect to $g$ is given by 
\begin{equation}
\delta_g\tilde{\mathcal E}=- \int_{0}^{T} \int_D\langle \ddot g+G\nabla_a q,h\rangle\id a \id t
\end{equation}
Since the first variation of $\tilde{\mathcal E}$ must vanish at its extremizers for any $h$ satisfying~\eqref{eq:h_bc}, we obtain 
\begin{equation}
\ddot g + G\nabla_a q = 0.
\label{eq:ddotg}
\end{equation}

Next, we define the scalar field $p$ through
\begin{equation}
p(t,g(t,a)):=q(t,a),\quad \forall a\in D. 
\label{eq:p}
\end{equation}
Differentiating with respect to $a$, we obtain
\begin{equation}
\nabla_a q (t,a) = [\nabla_a g(t,a)]^{\dagger}\nabla_x p(t,g(t,a)),
\end{equation}
or equivalently,
$$\nabla_x p(t,g(t,a))=[\nabla_a g(t,a)]^{-\dagger}\nabla_a q (t,a)= G(t,a)\nabla_a q(t,a).$$
This last identity, together with~\eqref{eq:ddotg}, yields the desired result
\begin{equation}
\ddot g(t,a) + \nabla_x p(t,g(t,a))=0.
\end{equation}
Note that the scalar field $p$ defined in~\eqref{eq:p} turns out to be the fluid pressure evaluated at
the point $g(t,a)$. In differential geometric terms, the Lagrange multiplier $q$ is the pullback of
the pressure $p$ under the map $g(t,\cdot)$.

\section{Conclusions}
Arnold~\cite{arnold1966} showed that the Euler equations of ideal incompressible fluids can be viewed as a geodesic flow on the manifold of volume preserving diffeomorphisms of the fluid domain. Despite the relevance of these results to fluid dynamicists, Arnold's proof and related follow-up works remain unaccessible to most of the fluid dynamics community because of their heavy use of differential geometric concepts and the properties of Lie groups.

Here we provided a simple derivation of Arnold's results by using classical concepts from calculus of variations.
Specifically, the Lagrangian flow map evolving the fluid configuration space from the initial to the final time, lies on a geodesic path minimizing the total kinetic energy of the fluid flow. The minimization is carried out with an additional constraint
that enforces the volume-preserving nature of the Lagrangian flow. The Lagrange multiplier 
enforcing this constraint coincides with the fluid pressure evaluated at the final time along each fluid trajectory (see equation~\eqref{eq:p}). 

Finally, Arnold~\cite{arnold1966} also interpreted Euler's equations for rigid body dynamics
as geodesic equations on an appropriately chosen manifold. We point out that this result can also be derived in
our constrained optimization framework using only classical notions of calculus of variations.

\begin{appendices}
\section{Proof of identity~\eqref{eq:G_id}}\label{app:G_id}
Using index notation and the convention of 
summation over repeated indices,
we have
\begin{equation}
\begin{aligned}
\nabla_a\cdot[G^\dagger h] & =\pard{}{a_i}[G_{ji}h_j]\\
& = \pard{G_{ji}}{a_i}h_j + G_{ji}\pard{h_j}{a_i}\nonumber\\
& = \langle\nabla_a\cdot G, h\rangle + \tr [G^\dagger\nabla_a h].
\end{aligned}
\end{equation}
\end{appendices}
\ \\

\textbf{Acknowledgements:} We are grateful to Prof. George Haller for his comments on an earlier draft of this paper. 


\end{document}